\documentclass[journal]{IEEEtran}
\usepackage{stfloats}
\usepackage{cite}
\usepackage{amsmath}
\usepackage{amsfonts}
\usepackage{dsfont}
\usepackage{graphicx}
\usepackage{changepage}
\usepackage{color}
\usepackage{amssymb}
\begin{document}

\title{Performance Analysis of Dual-Hop THz Transmission Systems over $\alpha$-$\mu$ Fading Channels\\ with Pointing Errors}

\author{Sai Li, Liang Yang 
\thanks{S. Li and L. Yang are with the College of Computer Science and Electronic Engineering, Hunan University, Changsha 410082,
China, (e-mail: lisa2019@hnu.edu.cn, liangy@hnu.edu.cn).}}
\maketitle

\begin{abstract}
In this paper, the performance of a dual-hop relaying terahertz (THz) wireless communication system is investigated. In particular, the behaviors of the two THz hops are determined by three factors, which are the deterministic path loss, the fading effects, and pointing errors. Assuming that both THz links are subject to the $\alpha$-$\mu$ fading with pointing errors, we derive exact expressions for the cumulative distribution function (CDF) and probability density function (PDF) of the end-to-end signal-to-noise ratio (SNR). Relying on the CDF and PDF, important performance metrics are evaluated, such as the outage probability, average bit error rate, and average channel capacity. Moreover, the asymptotic analyses are presented to obtain more insights. Results show that the dual-hop relaying scheme has better performance than the single THz link. The system's diversity order is $\min\left\{\frac{\phi_1}{2},\frac{\alpha_1\mu_1}{2},\phi_2,\alpha_2\mu_2\right\}$, where $\alpha_i$ and $\mu_i$ represent the fading parameters of the $i$-th THz link for $i\in(1,2)$, and $\phi_i$ denotes the pointing error parameter. In addition, we extend the analysis to a multi-relay cooperative system and derive the asymptotic symbol error rate expressions. Results demonstrate that the diversity order of the multi-relay system is $K\min\left\{\frac{\phi_1}{2},\frac{\alpha_1\mu_1}{2},\phi_2,\alpha_2\mu_2\right\}$, where $K$ is the number of relays. Finally, the derived analytical expressions are verified by Monte Carlo simulation.
\end{abstract}


\section{Introduction}
\IEEEPARstart{I}{nternet} of Things (IoT) is bound to explode with huge business requirements in the beyond fifth generation (B5G) era. With the increasing demand for wireless connection in IoT deployment, teraherhz (THz) is expected to become an attractive candidate spectrum for B5G networks due to the scarcity of spectrum resources. THz wireless transmission rate can exceed 100 Gbit/s, which provides an effective solution for ultra-short-distance and ultra-high-speed wireless transmission. THz waves represent electromagnetic waves with a frequency spectrum between 0.1 and 10 THz, which is between millimeter wave (mmWave) and infrared light \cite{1,2,3}. Compared with mmWave and optical wireless communication, THz has many unique advantages. For example, THz communications support both line-of-sight (LoS) and non-LoS (NLoS) propagation conditions, stronger anti-interference ability, better directionality, higher confidentiality, and can act as a reliable substitute in extreme weather conditions such as rain, fog, and dust \cite{4,5,6}.

However, THz band has not yet been fully developed, and still faces various challenges in different environments, like the high path loss and molecular absorption, channel characteristics, antenna misalignment, and hardware imperfections \cite{7,8}. A major issue refers to the high path loss of the THz frequency band in the urban environment, which implies THz communications having a limited propagation distance. To tackle this problem, appropriate path loss models applied to the THz propagation were presented in \cite{9,10,11,12,13}. For example, a ray-tracing approach for modeling short-distance THz band propagation channels was studied in \cite{10}. A THz band transmission model was proposed in \cite{11}, which considered the path loss and molecular absorption noise suffered by waves in the case of short-range propagation. In particular, J. Kokkoniemi \emph{et al.} proposed a simplified path attenuation model for 275-400 GHz frequency band \cite{12}. Subsequently, with the model proposed in [12], the performance of the THz system was studied in \cite{13}. Additionally, the impact of fading generated from THz propagation is another issue that cannot be neglected. In \cite{14}, the multipath fading of THz channels was modeled by Rician or Rayleigh or Nakagami-$m$ distributions. Furthermore, the shadowing effect has been experimentally verified in \cite{15}. Very recently, THz channel was modeled by an $\alpha$-$\mu$ fading distribution to accommodate the multipath effect \cite{7}. The $\alpha$-$\mu$ distribution is a generalized model, including several important distributions as special cases, such as Gamma, Nakagami-$m$, Weibull, and Rayleigh \cite{16}. By taking the $\alpha$-$\mu$ fading and misalignment effects into consideration, the analytical expressions of the outage probability (OP) and capacity for the THz link were derived and the performance under different fading conditions was discussed. What's more, the misalignment effects between the transmitter and the receiver antennas, also known as pointing errors, become a key issue in THz communications because they lead to significant performance degradation \cite{2,8}. Until now, the impact of pointing errors on the free space optical (FSO) link has been widely studied in \cite{17,18,19}. More recently, many studies have been made to study the effect of the pointing error on the THz-based network \cite{7,20,21}.

On the other hand, the cooperative diversity has been proposed to alleviate the fading caused by the transmission distance as well as multipath effects. The relaying scheme becomes a viable option for obtaining higher link capacity and wider coverage by dividing a long link with poor quality into two or more short links with better quality. At present, the relaying technique has been extensively developed in radio frequency (RF) and FSO communication systems. Meanwhile, a lot of efforts have been made to investigate the performance of mixed dual-hop RF-RF or RF-FSO transmission systems employing decode-and-forward (DF) or amplify-and-forward (AF) relaying (see e.g., \cite{22,23,24,25} and references therein). The mixed RF-FSO approach allows multiple RF messages to be aggregated into a single FSO link to achieve the maximum possible capacity. Motivated by this, mixed THz-RF relaying systems have been presented in order to enable several RF links to feed one high-rate THz link, thereby obtaining considerable performance. For instance, by employing the DF protocol, the outage and error performance of a THz-RF relaying system have been conducted in \cite{21}, where the THz and RF links experience $\alpha$-$\mu$ fading with pointing errors and Rayleigh fading, respectively. The authors in \cite{26,27} have investigated the performance of a mixed THz-RF system with the DF protocol, the exact expressions of OP and lower bound on ergodic capacity have been derived. In addition, considerable efforts have been devoted to evaluate the performance of dual-hop or multi-hop FSO transmission systems using the AF or DF relaying (see e.g., \cite{28},\cite{29} and references therein). In such system models, the relaying scheme can effectively mitigate the performance loss caused by fading and pointing errors compared to a single direct link. More recently, a dual-hop THz system was proposed in \cite{30} where both THz links experience the joint impacts of fading and misalignment effects. However, the authors in [30] only considered the DF case and analyzed the outage probability.

Similar to the model in \cite{30}, in this paper we comprehensively evaluate the performance of a dual-hop THz system with the fixed-gain relays. To the best of the authors' knowledge, the performance of dual-hop THz relaying systems with the fixed-gain AF protocol has not been studied in the literature yet. Specifically, the analytical expressions for the cumulative distribution function (CDF) and probability density function (PDF) of the end-to-end (e2e) signal-to-noise ratios (SNR) are derived. Relying on these obtained results, exact analytical expressions for the OP, average BER, and average channel capacity (ACC) of the considered system are derived in terms of the bivariate Fox's H-function (BFHF). To attain more useful insights, corresponding asymptotic results for the OP and average BER are investigated. Relying on the asymptotic results, we observe that the diversity order of the mixed dual-hop THz system depends on the fading parameters and pointing errors of both THz links. Results show that the performance of considered systems is better than the single THz link. Additionally, the performance of the mixed THz system deteriorates when the system under the conditions of far propagation distance and/or strong fading and/or strong pointing errors. Another contribution of this paper is that we extend the single-relay system to a multi-relay network, and present the asymptotic symbol error rate (SER) analysis for both relay selection and all-relay employed schemes.

The remainder of this paper is organized as follows. In Section II, the system and channel models are introduced. The tractable expressions of the CDF and PDF for a single THz link are given in this section. In Section III, the CDF and PDF of the e2e SNR for the considered dual-hop system are derived. In addition, we obtain the asymptotic CDF to gain more useful insights. In Section IV, we derive the exact analytical expressions of the OP, average BER, and ACC. The asymptotic OP and average BER are also derived. Moreover, the asymptotic SER analysis of the multi-relay case is presented in Section V. Section VI presents illustrative numerical results supplemented by Monte Carlo simulations to verify the accuracy of the performance metrics. Finally, insightful discussions are drawn in Section VII.

\section{System and Channel Models}
We consider a dual-hop THz communication system where a source (S) is communicating with a destination (D) through a single relay (R) with the fixed-gain AF protocol. We assume that the two THz links (i.e. S-R and R-D) follow the $\alpha$-$\mu$ fading with pointing errors. Moreover, S, R, and D are assumed to be equipped with a single highly directive antenna. In addition, we assume that an ideal RF front-end is employed, therefore the impact of hardware imperfections is neglected.

By using the fixed-gain AF relaying, the overall instantaneous SNR $\gamma_o$ of the dual-hop mixed THz system can be given by \cite{31,32}
\begin{align*}
\gamma_o=\frac{\gamma_1\gamma_2}{\gamma_2+C},
\tag{1}\label{1}
\end{align*}
where $C$ is a constant related to the amplification gain [31], and $\gamma_i$ is the instantaneous received SNR of the $i$th hop, $i\in(1, 2)$. From [7, Eq. (26)], the PDF of $\gamma_i$ can be derived by applying [21, Eq. (9)] as
\begin{align*}
f_{\gamma_i}(\gamma_i)=A_i\overline\gamma_{i}^{-\frac{\phi_i}{2}}  \gamma_{i}^{\frac{\phi_i}{2}-1}
\Gamma\left(\frac{\alpha_i\mu_i-\phi_i}{\alpha_i},B_i\left(\frac{\gamma_i}{\overline\gamma_i}\right)^{\frac{\alpha_i}{2}}\right),
\tag{2}\label{2}
\end{align*}
where $A_i=\frac{\phi_i\mu_i^{\frac{\phi_i}{\alpha_i}}h_{l,i}^{-\phi_i}}{2\hat{h}_{f,i}^{\alpha_i}A_{o,i}^{\phi_i}\Gamma(\mu_i)}$, $B_i=\frac{\mu_i}{(\hat{h}_{f,i}h_{l,i}A_{o,i})^{\alpha_{i}}}$, $\Gamma(\cdot)$ denotes the gamma function [33, Eq. (8.310)], $\Gamma(\cdot,\cdot)$ represents the incomplete gamma function [33, Eq. (8.350.2)], $\overline\gamma_i$ refers to the average SNR of the $i$th hop, $\alpha_i$ and $\mu_i$ stand for fading parameters of the $\alpha$-$\mu$ distribution, $\hat{h}_{f,i}$ holds for the $\alpha$-root mean value of the fading channel envelope, $A_{o,i}$ is the constant term that defines the pointing loss, and $\phi_i$ denotes the squared ratio between the equivalent beam radius and the pointing error displacement standard deviation $\sigma_{s,i}$ at the receiver [17, Eqs. (9) and (10)]. In addition, $h_{l,i}$ denotes the deterministic path loss of the $i$th THz channel which can be obtained as \cite{7,21}
\begin{align*}
h_{l,i}=\frac{c\sqrt{G_{t}G_{r}}}{4\pi fd_i}\exp \left(-\frac{1}{2}\beta(f)d_i\right),
\tag{3}\label{3}
\end{align*}
where $G_t$ and $G_r$ denote, respectively, the transmit and receive antenna gains of all nodes, $c$ refers to the speed of light, $f$ represents the operating frequency, $d_i$ represents the propagation distance of S-R and R-D links, and $\beta(f)$ stands for the absorption coefficient being function of the relative humidity $\varrho$, atmosphere pressure $p_a$, and temperature $T$ [7, Eqs. (8-17)].

Based on (2), the PDF of $\gamma_i$ can be rewritten by employing [34, Eq. (8.4.16/2)] as
\begin{align*}
f_{\gamma_i}(\gamma_i)=A_i&\overline\gamma_{i}^{-\frac{\phi_i}{2}}\gamma_{i}^{\frac{\phi_i}{2}-1}\\ &\times
\, {\mathrm{G}}_{1,2}^{2,0}\left [{{B_i\left(\frac{\gamma_i}{\overline\gamma_i}\right)^{\frac{\alpha_i}{2}}}\left |{ \begin{matrix} {1}
\\ {0,\frac{\alpha_i\mu_i-\phi_i}{\alpha_i}} \\ \end{matrix} }\right . }\right ]\!,
\tag{4}\label{4}
\end{align*}
where ${\mathrm{G}}_{p,q}^{m,n}[\cdot]$ is the Meijer's G-function [33, Eq. (9.301)]. By using the primary definition of Meijer's G-function, the CDF of the THz link can be derived as
\begin{align*}
F_{\gamma_i}(\gamma_i)=&\frac{2A_i\overline\gamma_{i}^{-\frac{\phi_i}{2}}  \gamma_{i}^{\frac{\phi_i}{2}}}{\alpha_i}\\ &\times
\, {\mathrm{G}}_{2,3}^{2,1}\left [{{B_i\left(\frac{\gamma_i}{\overline\gamma_i}\right)^{\frac{\alpha_i}{2}}}
\left |{ \begin{matrix} {1{-}\frac{\phi_i}{\alpha_i},1}
\\ {0,\frac{\alpha_i\mu_i-\phi_i}{\alpha_i},-\frac{\phi_i}{\alpha_i}} \\ \end{matrix} }\right . }\right ]\!.
\tag{5}\label{5}
\end{align*}

\section{Statistical Analysis}
In this section, exact analytical expressions for the CDF and PDF of the e2e SNR are derived.
In addition, we derive the asymptotic CDF at high SNR regimes to get more useful insights.
\subsection{Cumulative Distribution Function}
\subsubsection{Exact Analysis}
From \cite{35,36}, the CDF of the e2e SNR $\gamma_o$ is obtained by taking a series of transformations as
\begin{align*}
F_{\gamma_o}(\gamma)=&\int_{0}^{\infty}
P\left[\frac{\gamma_1\gamma_2}{\gamma_2+C}
<\gamma \Bigg|\gamma_1\right]f_{\gamma_1}(\gamma_1)d\gamma_1 \\
=&F_{\gamma_1}(\gamma)+\int_{0}^{\infty}F_{\gamma_2}
\left(\frac{C\gamma}{x}\right)f_{\gamma_1}(x+\gamma)dx.
\tag{6}\label{6}
\end{align*}
By substituting (4) and (5) into (6), the CDF can be obtained as
\begin{align*}
&F_{\gamma_o}(\gamma)=\frac{2A_1\overline\gamma_{1}^{-\frac{\phi_1}{2}}  \gamma^{\frac{\phi_1}{2}}}{\alpha_1}
\, {\mathrm{G}}_{2,3}^{2,1}\left [{{B_1\left(\frac{\gamma}{\overline\gamma_1}\right)^{\frac{\alpha_1}{2}}}
\left |{ \begin{matrix} {1{-}\frac{\phi_1}{\alpha_1},1}
\\ {0,\frac{\alpha_1\mu_1-\phi_1}{\alpha_1},-\frac{\phi_1}{\alpha_1}} \\ \end{matrix} }\right . }\right ]\!\\& +\frac{2A_1A_2\overline\gamma_{1}^{-\frac{\phi_1}{2}}\overline\gamma_{2}^{-\frac{\phi_2}{2}}
C^{\frac{\phi_2}{2}}\gamma^{\frac{\phi_1}{2}}}{\alpha_2}{\rm {H}_{1,0:4,2:2,2}^{0,1:1,3:0,2}}\\
&\left [{\!\!\left .{ \begin{matrix} \left({1{+}\frac{\phi_1}{2}{-}\frac{\phi_2}{2};-\frac{\alpha_2}{2},\frac{\alpha_1}{2}}\right)
\\ -\\ (1,1) \left(1{-}\mu_2{+}\frac{\phi_2}{\alpha_2},1\right) \left(\frac{\phi_2}{2},\frac{\alpha_2}{2}\right)
\left(1{+}\frac{\phi_2}{\alpha_2},1\right) \\
\left(\frac{\phi_2}{\alpha_2},1\right) (0,1)\\ (1,1) \left(1{-}\mu_1{+}\frac{\phi_1}{\alpha_1},1\right) \\
(0,1) \left(\frac{\phi_1}{2},\frac{\alpha_1}{2}\right) \end{matrix} }\right |\!
\frac{\overline\gamma_2^{\frac{\alpha_2}{2}}}{B_2C^{\frac{\alpha_2}{2}}}, \!\frac{\overline\gamma_{1}^{\frac{\alpha_1}{2}}}
{B_1\gamma^{\frac{\alpha_1}{2}}} \!\!}\right ],
\tag{7}\label{7}
\end{align*}
where ${\mathrm{H}}_{p1,q1:p2,q2:p3,q3}^{0,n1:m2,n2:m3,n3}[\cdot,\cdot]$ denotes the BFHF [37, Eq. (2.57)]. It is worth noting that the BFHF can be effectively evaluated in MATLAB or MATHEMATICA, and the available code implementations were elaborated in \cite{38} and \cite{39}.

\textit{Proof:} See Appendix A.

\emph{Special Case:} Consider $\alpha_1=\alpha_2=2$ corresponding to both THz links suffering from the Nakagami-$m$ fading. Then, we have
\begin{align*}
&F_{\gamma_o}^{N}(\gamma)=\zeta_1\overline\gamma_{1}^{-\frac{\phi_1}{2}} \gamma^{\frac{\phi_1}{2}}
\, {\mathrm{G}}_{2,3}^{2,1}\left [{{\frac{\zeta_2\gamma}{\overline\gamma_1}}
\left |{ \begin{matrix} {1{-}\frac{\phi_1}{2},1}
\\ {0,\mu_1{-}\frac{\phi_1}{2},-\frac{\phi_1}{2}} \\ \end{matrix} }\right . }\right ]\!\\& +\zeta_1\zeta_3\overline\gamma_{1}^{-\frac{\phi_1}{2}}\overline\gamma_{2}^{-\frac{\phi_2}{2}}
C^{\frac{\phi_2}{2}}\gamma^{\frac{\phi_1}{2}}{\rm {H}_{1,0:4,2:2,2}^{0,1:1,3:0,2}}\\
&\left [{\!\!\left .{ \begin{matrix} \left({1{+}\frac{\phi_1}{2}{-}\frac{\phi_2}{2};-1,1}\right)
\\ -\\ (1,1) \left(1{-}\mu_2{+}\frac{\phi_2}{2},1\right) \left(\frac{\phi_2}{2},1\right)
\left(1{+}\frac{\phi_2}{2},1\right) \\
\left(\frac{\phi_2}{2},1\right) (0,1)\\ (1,1) \left(1{-}\mu_1{+}\frac{\phi_1}{2},1\right) \\
(0,1) \left(\frac{\phi_1}{2},1\right) \end{matrix} }\right |\!
\frac{\overline\gamma_2}{\zeta_4C}, \!\frac{\overline\gamma_{1}}{\zeta_2\gamma} \!\!}\right ],
\tag{8}\label{8}
\end{align*}
where $\zeta_1=\frac{\phi_1\mu_1^{\frac{\phi_1}{2}}h_{l,1}^{-\phi_1}}{2\hat{h}_{f,1}^{2}A_{o,1}^{\phi_1}\Gamma(\mu_1)}$,
$\zeta_2=\frac{\mu_1}{(\hat{h}_{f,1}h_{l,1}A_{o,1})^{2}}$, $\zeta_3=\frac{\phi_2\mu_2^{\frac{\phi_2}{2}}h_{l,2}^{-\phi_2}}{2\hat{h}_{f,2}^{2}A_{o,2}^{\phi_2}\Gamma(\mu_2)}$,
$\zeta_4=\frac{\mu_2}{(\hat{h}_{f,2}h_{l,2}A_{o,2})^{2}}$. Furthermore, (8) can be further simplified to the CDF that both links suffer from Rayleigh fading by setting $\mu_1=\mu_2=1$. Please note that these results are also novel and have not been presented in the literature yet.

\subsubsection{Asymptotic Analysis}
Since the analytical expression of the CDF given by (7) is a complex expression in terms of the BFHF, it can only provide limited physical insights. As such, the asymptotic analysis of the CDF at high SNRs is also developed. By letting $\overline\gamma_1\to \infty$ and $\overline\gamma_2\to \infty$, applying [40, Eq. (07.34.06.0040.01)] and [41, Eq. (1.8.4)], and doing some algebraic operations, the asymptotic CDF can be derived as (9), shown at the top of the next page. Therefore, one can see that the CDF is reduced to a very simple form, which can be used to get further insights.

\textit{Proof:} See Appendix B.

\subsection{Probability Density Function}
For the fixed-gain AF case, the PDF can be expressed by calculating the derivative of (6) with respect to $\gamma$ as [36, Eq. (51)]
\begin{align*}
f_{\gamma_{o}}(\gamma)=\int_{\gamma}^{\infty}
\frac{C\gamma_{1}}{(\gamma_{1}-\gamma)^2}
f_{\gamma_{2}}\left(\frac{C\gamma}{\gamma_{1}-\gamma}\right)
f_{\gamma_{1}}(\gamma_{1})d\gamma_{1}.
\tag{10}\label{10}
\end{align*}
By inserting (4) into (10) and doing some algebraic operations, the PDF of the dual-hop THz system is attained as
\begin{align*}
&f_{\gamma_o}(\gamma)=A_1A_2\overline\gamma_{1}^{-\frac{\phi_1}{2}}\overline\gamma_{2}^{-\frac{\phi_2}{2}}
C^{\frac{\phi_2}{2}}\gamma^{\frac{\phi_1}{2}-1}{\rm {H}_{1,0:3,1:2,2}^{0,1:0,3:0,2}}\\
&\left [{\!\!\left .{ \begin{matrix} \left({1{+}\frac{\phi_1}{2}{-}\frac{\phi_2}{2};-\frac{\alpha_2}{2},\frac{\alpha_1}{2}}\right)
\\ -\\ (1,1) \left(1{-}\mu_2{+}\frac{\phi_2}{\alpha_2},1\right) \left(1{+}\frac{\phi_2}{2},\frac{\alpha_2}{2}\right) \\
(0,1)\\ (1,1) \left(1{-}\mu_1{+}\frac{\phi_1}{\alpha_1},1\right) \\
(0,1) \left(1{+}\frac{\phi_1}{2},\frac{\alpha_1}{2}\right) \end{matrix} }\right |\!
\frac{\overline\gamma_2^{\frac{\alpha_2}{2}}}{B_2C^{\frac{\alpha_2}{2}}}, \!\frac{\overline\gamma_{1}^{\frac{\alpha_1}{2}}}
{B_1\gamma^{\frac{\alpha_1}{2}}} \!\!}\right ].
\tag{11}\label{11}
\end{align*}

\textit{Proof:} See Appendix C.

\begin{figure*}[t]
\begin{align*}
F_{\gamma_o}^{A}(\gamma)& \underset{\overline\gamma_{1},~\overline\gamma_{2} \to \infty}\approx \frac{2A_1\Gamma\left(\frac{\alpha_1\mu_1{-}\phi_1}{\alpha_1}\right)\Gamma\left(\frac{\phi_1}{\alpha_1}\right)}
{\alpha_1\Gamma\left(1{+}\frac{\phi_1}{\alpha_1}\right)}
\left(\frac{\gamma}{\overline\gamma_{1}}\right)^{\frac{\phi_1}{2}}
{+}\frac{2A_1B_{1}^{\frac{\alpha_1\mu_1{-}\phi_1}{\alpha_1}}\Gamma\left(-\frac{\alpha_1\mu_1{-}\phi_1}
{\alpha_1}\right)}{\alpha_1\mu_1\Gamma\left(1{-}\frac{\alpha_1\mu_1{-}\phi_1}{\alpha_1}\right)}
\left(\frac{\gamma}{\overline\gamma_{1}}\right)^{\frac{\alpha_1\mu_1}{2}}\\
&+\frac{4A_1A_2B_{1}^{\frac{\phi_2{-}\phi_1}{\alpha_1}}\Gamma\left(\frac{\alpha_2\mu_2{-}\phi_2}{\alpha_2}\right)
\Gamma\left(\frac{\phi_1{-}\phi_2}{\alpha_1}\right)\Gamma\left(\mu_1\frac{\phi_2}{\alpha_1}\right)\Gamma\left(\frac{\phi_2}{2}\right)}
{\alpha_1\alpha_2\Gamma\left(1{-}\frac{\phi_2{-}\phi_1}{\alpha_1}\right)
\Gamma\left(1{+}\frac{\phi_2}{\alpha_2}\right)}\left(\frac{C\gamma}{\overline\gamma_{1}\overline\gamma_{2}}\right)^{\frac{\phi_2}{2}}\\
&+\frac{4A_1A_2B_{1}^{\frac{\alpha_2\mu_2{-}\phi_1}{\alpha_1}}B_{2}^{\frac{\alpha_2\mu_2{-}\phi_2}{\alpha_2}}
\Gamma\left(-\frac{\alpha_2\mu_2{-}\phi_2}{\alpha_2}\right)
\Gamma\left(\frac{\phi_1{-}\alpha_2\mu_2}{\alpha_2}\right)\Gamma\left(\mu_1{-}\frac{\alpha_2\mu_2}{\alpha_1}\right)
\Gamma(\mu_2)}{\alpha_1\alpha_2\Gamma\left(1{-}\frac{\alpha_2\mu_2{-}\phi_2}
{\alpha_2}\right)\Gamma\left(1{+}\frac{\phi_1}{\alpha_1}-\frac{\alpha_2\mu_2}{\alpha_1}\right)\Gamma(1{+}\mu_2)}
\left(\frac{C\gamma}{\overline\gamma_{1}\overline\gamma_{2}}\right)^{\frac{\alpha_2\mu_2}{2}}\\
&+\frac{4A_1A_2B_{2}^{\frac{\phi_1{-}\phi_2}{\alpha_2}}\Gamma\left(-\frac{\phi_1{-}\phi_2}{\alpha_2}\right)
\Gamma\left(\mu_2{-}\frac{\phi_1}{\alpha_2}\right)\Gamma\left(\mu_1{-}\frac{\phi_1}{\alpha_1}\right)
\Gamma\left(\frac{\phi_1}{\alpha_2}\right)}
{\alpha_{2}^2\Gamma\left(1{-}\frac{\phi_1{-}\phi_2}{\alpha_2}\right)
\Gamma\left(1{-}\frac{\phi_2{-}\phi_1}{\alpha_1}{-}\frac{\phi_1{-}\phi_2}{\alpha_2}\right)
\Gamma\left(1{+}\frac{\phi_1}{\alpha_2}\right)}
\left(\frac{C\gamma}{\overline\gamma_{1}\overline\gamma_{2}}\right)^{\frac{\phi_1}{2}}\\
&+\frac{4A_1A_2B_{1}^{\frac{\alpha_1\mu_1{-}\phi_1}{\alpha_1}}B_{2}^{\frac{\alpha_1\mu_1{-}\phi_2}{\alpha_2}}
\Gamma\left(-\frac{\alpha_1\mu_1{-}\phi_2}{\alpha_2}\right)\Gamma\left(\mu_2{-}\frac{\alpha_1\mu_1}{\alpha_2}\right)
\Gamma\left(\frac{\phi_1}{\alpha_1}{-}\mu_1\right)\Gamma\left(\frac{\alpha_1\mu_1}{\alpha_2}\right)}
{\alpha_{2}^{2}\Gamma\left(1{-}\frac{\alpha_1\mu_1{-}\phi_2}{\alpha_2}\right)
\Gamma\left(1{+}\frac{\phi_1}{\alpha_1}{-}\mu_1\right)
\Gamma\left(1{+}\frac{\alpha_1\mu_1}{\alpha_2}\right)}
\left(\frac{C\gamma}{\overline\gamma_{1}\overline\gamma_{2}}\right)^{\frac{\alpha_1\mu_1}{2}}.
\tag{9}\label{9}
\end{align*}
\hrulefill
\end{figure*}

\section{Performance Analysis}
In this section, important performance metrics of the dual-hop THz system are evaluated, namely the OP, average BER, and ACC. In addition, we provide the asymptotic analysis of the OP and average BER to obtain the considered system's diversity order.

\subsection{Outage Probability}
\subsubsection{Exact Analysis}
In general, if the received instantaneous SNR is lower than a certain received SNR $\gamma_{th}$, the dual-hop relaying system undergoes outage. The OP of the considered AF relaying system can be mathematically written by using (7) as $P_{out}=\Pr[\gamma_o <\gamma_{th}]=F_{\gamma_o}(\gamma_{th})$.

\subsubsection{Asymptotic Analysis}
At high SNRs, the asymptotic OP can be obtained directly from (9), that is, $P_{out}\to F_{\gamma_o}^{A}(\gamma_{th})$. From \cite{42}, assuming $\overline\gamma_i\to \infty$, a very simple form of the OP is asymptotically proposed as $P_{out}\approx (G_c\cdot \overline\gamma_i)^{-G_d}$, where $G_c$ refers to the coding gain, $G_d$ stands for the diversity gain. After a careful inspection, from the asymptotic result of the OP, we can notice that the diversity order of the considered system can be easily deduced as
\begin{align*}
G_d = \min\left\{\frac{\phi_1}{2},\frac{\alpha_1\mu_1}{2},\phi_2,\alpha_2\mu_2\right\}.
\tag{12}\label{12}
\end{align*}
As can be seen from (12), it is easily noticed that the diversity gain of the dual-hop THz relaying system depends on the multipath fading parameters (i.e. $\alpha_1$, $\mu_1$, $\alpha_2$, and $\mu_2$) and parameters related to pointing errors (i.e. $\phi_1$ and $\phi_2$) of both S-R and R-D links. As far as the authors are aware, this remark has not been presented in the literature before.

\subsection{Average Bit Error Rate}
\subsubsection{Exact Analysis}
From[22, Eq. (25)], the average BER for various modulation schemes is given by
\begin{align*}
\overline{P}_e = \frac{q^p}{2\Gamma(p)}\int_{0}^{\infty}\gamma^{p-1}e^{-q\gamma}F_{\gamma_o}(\gamma)d\gamma,
\tag{13}\label{13}
\end{align*}
where the different values of $p$ and $q$ hold for various modulation methods, $\{p=0.5,q=1\}$ and $\{p=1,q=1\}$ are used for binary phase shift keying (BPSK) and differential PSK (DPSK), respectively. Therefore, substituting (7) into (13), employing [40, Eqs. (07.34.03.0228.01) and (07.34.21.0012.01)] and [33, Eq. (3.326.2)] yields
\begin{align*}
&\overline{P}_{e}=\frac{A_1(q\overline\gamma_{1})^{-\frac{\phi_1}{2}}}{\Gamma(p)\alpha_1}\\ &\times
\, {\mathrm{H}}_{3,3}^{2,2}\left [{{\frac{B_1}{(q\overline\gamma_1)^{\frac{\alpha_1}{2}}}}
\left |{ \begin{matrix} {\left(1{-}p{-}\frac{\phi_1}{\alpha_1},\frac{\alpha_1}{2}\right) \left(1{-}\frac{\phi_1}{\alpha_1},1\right) (1,1)}
\\ {(0,1) \left(\frac{\alpha_1\mu_1-\phi_1}{\alpha_1},1\right) \left(-\frac{\phi_1}{\alpha_1},1\right)} \\ \end{matrix} }\right . }\right ]\!\\& +\frac{A_1A_2\overline\gamma_{1}^{-\frac{\phi_1}{2}}\overline\gamma_{2}^{-\frac{\phi_2}{2}}
C^{\frac{\phi_2}{2}}q^{-\frac{\phi_1}{2}}}{\alpha_2\Gamma(p)}{\rm {H}_{1,0:4,2:2,3}^{0,1:1,3:1,2}}\\
&\left [{\!\!\left .{ \begin{matrix} \left({1{+}\frac{\phi_1}{2}{-}\frac{\phi_2}{2};-\frac{\alpha_2}{2},\frac{\alpha_1}{2}}\right)
\\ -\\ \epsilon_2 \\
\left(\frac{\phi_2}{\alpha_2},1\right) (0,1)\\ (1,1) \left(1{-}\mu_1{+}\frac{\phi_1}{\alpha_1},1\right) \\
\left(\frac{\phi_1}{2}{+}p,\frac{\alpha_1}{2}\right) (0,1) \left(\frac{\phi_1}{2},\frac{\alpha_1}{2}\right)  \end{matrix} }\right |\!
\frac{\overline\gamma_2^{\frac{\alpha_2}{2}}}{B_2C^{\frac{\alpha_2}{2}}}, \!\frac{(q\overline\gamma_{1})^{\frac{\alpha_1}{2}}}
{B_1} \!\!}\right ],
\tag{14}\label{14}
\end{align*}
where $\epsilon_2=\left\{(1,1) \left(1{-}\mu_2{+}\frac{\phi_2}{\alpha_2},1\right) \left(\frac{\phi_2}{2},\frac{\alpha_2}{2}\right)
\left(1{+}\frac{\phi_2}{\alpha_2},1\right)\right\}$, ${\mathrm{H}}_{p,q}^{m,n}[\cdot]$ is the Fox's H-function [37, Eq. (1.2)]. Please note that the Fox's H-function can be effectively calculated, and the MATHEMATICA implementation code has been provided in \cite{43}.

\textit{Proof:} See Appendix D.

\begin{figure*}[t]
\begin{align*}
\overline{P}_{e}^{A} &\underset{\overline\gamma_{1},~\overline\gamma_{2} \to \infty}\approx \frac{A_1\Gamma\left(\frac{\alpha_1\mu_1{-}\phi_1}{\alpha_1}\right)\Gamma\left(\frac{\phi_1}{\alpha_1}\right)
\Gamma\left(p{+}\frac{\phi_1}{2}\right)}{\alpha_1\Gamma\left(1{+}\frac{\phi_1}{\alpha_1}\right)\Gamma(p)}
\left(\frac{1}{q\overline\gamma_{1}}\right)^{\frac{\phi_1}{2}}+\frac{A_1B_{1}^{\frac{\alpha_1\mu_1{-}\phi_1}{\alpha_1}}
\Gamma\left(-\frac{\alpha_1\mu_1{-}\phi_1}{\alpha_1}\right)\Gamma\left(p{+}\frac{\alpha_1\mu_1}{2}\right)}
{\alpha_1\mu_1\Gamma\left(1{-}\frac{\alpha_1\mu_1{-}\phi_1}{\alpha_1}\right)\Gamma(p)}
\left(\frac{1}{q\overline\gamma_{1}}\right)^{\frac{\alpha_1\mu_1}{2}}\\
&+\frac{2A_1A_2B_{1}^{\frac{\phi_2{-}\phi_1}{\alpha_1}}\Gamma\left(\frac{\alpha_2\mu_2{-}\phi_2}{\alpha_2}\right)
\Gamma\left(\frac{\phi_1{-}\phi_2}{\alpha_1}\right)\Gamma\left(\mu_1{-}\frac{\phi_2}{\alpha_1}\right)
\Gamma\left(\frac{\phi_2}{\alpha_2}\right)\Gamma\left(p{+}\frac{\phi_2}{2}\right)}
{\alpha_1\alpha_2\Gamma\left(1{-}\frac{\phi_2{-}\phi_1}{\alpha_1}\right)
\Gamma\left(1{+}\frac{\phi_2}{\alpha_2}\right)\Gamma(p)}
\left(\frac{C}{q\overline\gamma_{1}\overline\gamma_{2}}\right)^{\frac{\phi_2}{2}}\\
&+\frac{2A_1A_2B_{1}^{\frac{\alpha_2\mu_2{-}\phi_1}{\alpha_1}}B_{2}^{\frac{\alpha_2\mu_2{-}\phi_2}{\alpha_2}}
\Gamma\left(-\frac{\alpha_2\mu_2{-}\phi_2}{\alpha_2}\right)
\Gamma\left(\frac{\phi_1{-}\alpha_2\mu_2}{\alpha_2}\right)\Gamma\left(\mu_1{-}\frac{\alpha_2\mu_2}{\alpha_1}\right)
\Gamma(\mu_2)\Gamma\left(p{+}\frac{\alpha_2\mu_2}{2}\right)}
{\alpha_1\alpha_2\Gamma\left(1{-}\frac{\alpha_2\mu_2{-}\phi_2}
{\alpha_2}\right)\Gamma\left(1{+}\frac{\phi_1}{\alpha_1}-\frac{\alpha_2\mu_2}{\alpha_1}\right)\Gamma(1{+}\mu_2)\Gamma(p)}
\left(\frac{C}{q\overline\gamma_{1}\overline\gamma_{2}}\right)^{\frac{\alpha_2\mu_2}{2}}\\
&+\frac{2A_1A_2B_{2}^{\frac{\phi_1{-}\phi_2}{\alpha_2}}\Gamma\left(-\frac{\phi_1{-}\phi_2}{\alpha_2}\right)
\Gamma\left(\mu_2{-}\frac{\phi_1}{\alpha_2}\right)\Gamma\left(\mu_1{-}\frac{\phi_1}{\alpha_1}\right)
\Gamma\left(\frac{\phi_1}{\alpha_2}\right)\Gamma\left(p{+}\frac{\phi_1}{2}\right)}
{\alpha_{2}^{2}\Gamma\left(1{-}\frac{\phi_1{-}\phi_2}{\alpha_2}\right)
\Gamma\left(1{-}\frac{\phi_2{-}\phi_1}{\alpha_1}{-}\frac{\phi_1{-}\phi_2}{\alpha_2}\right)
\Gamma\left(1{+}\frac{\phi_1}{\alpha_2}\right)\Gamma(p)}
\left(\frac{C}{q\overline\gamma_{1}\overline\gamma_{2}}\right)^{\frac{\phi_1}{2}}\\
&+\frac{2A_1A_2B_{1}^{\frac{\alpha_1\mu_1{-}\phi_1}{\alpha_1}}B_{2}^{\frac{\alpha_1\mu_1{-}\phi_2}{\alpha_2}}
\Gamma\left(-\frac{\alpha_1\mu_1{-}\phi_2}{\alpha_2}\right)\Gamma\left(\mu_2{-}\frac{\alpha_1\mu_1}{\alpha_2}\right)
\Gamma\left(\frac{\phi_1}{\alpha_1}{-}\mu_1\right)\Gamma\left(\frac{\alpha_1\mu_1}{\alpha_2}\right)
\Gamma\left(p{+}\frac{\alpha_1\mu_1}{2}\right)}
{\alpha_{2}^{2}\Gamma\left(1{-}\frac{\alpha_1\mu_1{-}\phi_2}{\alpha_2}\right)
\Gamma\left(1{+}\frac{\phi_1}{\alpha_1}{-}\mu_1\right)
\Gamma\left(1{+}\frac{\alpha_1\mu_1}{\alpha_2}\right)\Gamma(p)}
\left(\frac{C}{q\overline\gamma_{1}\overline\gamma_{2}}\right)^{\frac{\alpha_1\mu_1}{2}}.
\tag{15}\label{15}
\end{align*}
\hrulefill
\end{figure*}

\subsubsection{Asymptotic Analysis}
The exact expression of the average BER is also a complex expression in terms of the BFHF. As such, by inserting (9) into (13) and employing [33, Eq. (3.326.2)], the asymptotic average BER can be obtained as (15), shown at the top of the next page. As a double check, from the asymptotic result of the average BER, one can again conclude that the system diversity order is $G_d=\min\left\{\frac{\phi_1}{2},\frac{\alpha_1\mu_1}{2},\phi_2,\alpha_2\mu_2\right\}$, which is equal to the result previously obtained in the OP analysis.

\subsection{Average Channel Capacity}
From \cite{22,28}, the ACC can be formulated as
\begin{align*}
&\overline{C}=\frac{1}{2\ln(2)}\int_{0}^{\infty} \ln(1+\gamma)f_{\gamma_o}(\gamma)d\gamma.
\tag{16}\label{16}
\end{align*}
Substituting (11) into (16), converting $\ln (1+\gamma)$ into the representation of the Meijer's G-function [34, Eq. (8.4.6/5)] and then following [37, Eqs. (2.9) and (2.57)], we have
\begin{align*}
&\overline{C}=\frac{A_1A_2\overline\gamma_{1}^{-\frac{\phi_1}{2}}\overline\gamma_{2}^{-\frac{\phi_2}{2}}
C^{\frac{\phi_2}{2}}}{2\ln(2)}{\rm {H}_{1,0:3,1:3,3}^{0,1:0,3:1,3}}\\
&\left [{\!\!\left .{ \begin{matrix} \left({1{+}\frac{\phi_1}{2}{-}\frac{\phi_2}{2};-\frac{\alpha_2}{2},\frac{\alpha_1}{2}}\right)
\\ -\\ (1,1) \left(1{-}\mu_2{+}\frac{\phi_2}{\alpha_2},1\right) \left(1{+}\frac{\phi_2}{2},\frac{\alpha_2}{2}\right) \\
(0,1)\\ (1,1) \left(1{-}\mu_1{+}\frac{\phi_1}{\alpha_1},1\right) \left(1{+}\frac{\phi_2}{2},\frac{\alpha_1}{2}\right)\\
\left(1{+}\frac{\phi_1}{2},\frac{\alpha_1}{2}\right) (0,1) \left(\frac{\phi_1}{2},\frac{\alpha_1}{2}\right) \end{matrix} }\right |\!
\frac{\overline\gamma_2^{\frac{\alpha_2}{2}}}{B_2C^{\frac{\alpha_2}{2}}}, \!\frac{\overline\gamma_{1}^{\frac{\alpha_1}{2}}}{B_1} \!\!}\right ].
\tag{17}\label{17}
\end{align*}

\section{Extension to Multi-Relay Systems}
\subsection{System Models}
Previously results are only applicable to the single relay system. In this section, we consider a more general multi-relay cooperative system, as shown in Fig. 1. It is worth noting that, unlike the RF network, the THz link radiates a narrow beam with concentrated energy in a smaller area. Therefore, we assume that $K$ transmitters are needed to completed signal transmission. More specifically, the signal at $\mathrm{S}_{i}, i\in\{1,...,K\}$ is correspondingly transmitted to the relay $\mathrm{R}_{i}$, and then the signal at $\mathrm{R}_{i}$ is amplified and retransmitted to D. Notice that the fixed-gain AF protocol is still employed in this section. We assume that the channels between the relays are orthogonal with each other over time. In addition, both best-relay-selection (BRS) and conventional all-relay participating (ARP) schemes are considered. Subsequently, the asymptotic SER expressions of two schemes are investigated and the performance comparison is provided. For fair comparison, we assume that BRS and ARP schemes have the same total power.

For the ARP scheme, by using maximum ratio combining, the received signals from all branches are weighted and then combined to maximize the output SNR. Therefore, the received SNR at D is [44][45]
\begin{align*}
\gamma_{o}^{ARP} = \frac{1}{2K}\sum_{i=1}^{K}\gamma_{o,i},
\tag{18}\label{18}
\end{align*}
where the factor $1/(2K)$ denotes that $K$ sources and $K$ relays use the $1/(2K)$ transmit power, respectively, $\gamma_{o,i}=\frac{\gamma_{1,i}\gamma_{i,2}}{\gamma_{i,2}+C}$ is the e2e SNR of S-$\mathrm{R}_{i}$-D link, where $\gamma_{1,i}$ and $\gamma_{i,2}$ stand for the instantaneous SNRs of S-$\mathrm{R}_{i}$ and $\mathrm{R}_{i}$-D links, respectively.

While for the BRS scheme, the output SNR can be expressed as [45, Eq. (2)]
\begin{align*}
\gamma_{o}^{BRS} = \frac{1}{2}\operatorname*{max}\limits_{i} \gamma_{o,i},
\tag{19}\label{19}
\end{align*}
where the factor $1/2$ represents that the selected $\mathrm{S}_{i}$ and $\mathrm{R}_{i}$ are set as $1/2$ of the transmit power, respectively. Next, using previously obtained statistics, we derive the asymptotic SER expressions of the multi-relay systems. From [46], the SER with $M$-ary PSK modulation is given by
\begin{align*}
P_{e}^{Q}= \frac{1}{\pi}\int_{0}^{\pi-\frac{\pi}{M}}M_{\gamma_{o}^{Q}}\left(\frac{\sin^2(\pi/M)}{\sin^2\theta}\right)d\theta,
\tag{20}\label{20}
\end{align*}
where $Q\in \{ARP,BRS\}$. For $M=2$, the above expression can be equivalent to the SER of BPSK modulation.

\begin{figure}[t]
    \centering
    \includegraphics[width=3.5in]{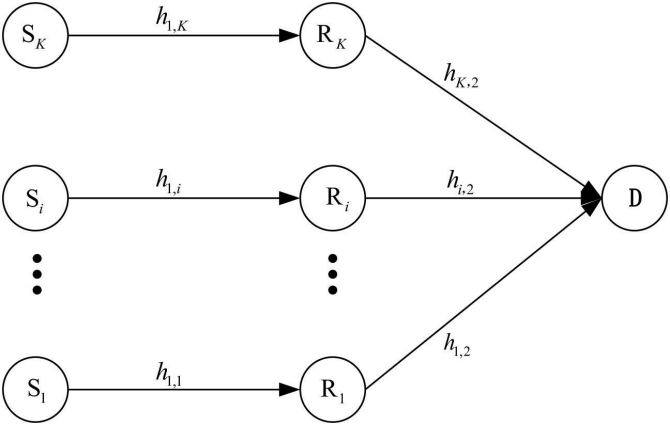}
    \caption{Multi-relay cooperative system model.}
\end{figure}

\subsection{ARP scheme}
From (9), assuming that $\overline\gamma_1=\overline\gamma_2=\overline\gamma$, the CDF of the single relay system can be asymptotically rewritten as
\begin{align*}
F_{\gamma_{o,i}}(\gamma)\to \frac{D\gamma^{v}}{\overline\gamma^{G_d}},
\tag{21}\label{21}
\end{align*}
where $D$ is the constant, $v=\min\{\frac{\phi_1}{2},\frac{\alpha_1\mu_1}{2},\frac{\phi_2}{2},\frac{\alpha_2\mu_2}{2}\}$. The moment generating function (MGF) of $\gamma_{o,i}$ can be formulated by employing the definition $M_{\gamma}(s)\triangleq s\int_{0}^{\infty} e^{-\gamma s}F_{\gamma}(\gamma) d\gamma$ as [28, Eq. (12)]
\begin{align*}
M_{\gamma_{o,i}}(s)=\frac{D\Gamma(v+1)}{\overline\gamma^{G_d}}s^{-v}.
\tag{22}\label{22}
\end{align*}
From [47, Eq. (7)], the MGF of $\gamma_{o}^{ARP}$ is expressed as
\begin{align*}
M_{\gamma_{o}^{ARP}}(s)=\left[M_{\gamma_{o,i}}\left(\frac{s}{2K}\right)\right]^K.
\tag{23}\label{23}
\end{align*}
By substituting (23) into (20) and assuming that $M=2$, the asymptotic SER of the ARP scheme is derived as
\begin{align*}
P_{e}^{ARP}\to \frac{D^K(\Gamma(v+1))^K\Gamma(Kv+\frac{1}{2})(2K)^{Kv}}{2\sqrt{\pi}\overline\gamma^{KG_d}\Gamma(Kv+1)}.
\tag{24}\label{24}
\end{align*}
As can be seen from (24), it reveals that the diversity order of the ARP system is $G_{d}^{ARP}=K\min\left\{\frac{\phi_1}{2},\frac{\alpha_1\mu_1}{2},\phi_2,\alpha_2\mu_2\right\}$, which depends on the number of relays and fading parameters.

\subsection{BRS scheme}
For the BRS scheme, the CDF of $\gamma_{o}^{BRS}$ can be written as $F_{\gamma_{o}^{BRS}}(\gamma)=[F_{\gamma_{o,i}}(\gamma)]^{K}$ [47, Eq. (13)]. With aid of (19) and (21), the asymptotic CDF of the BRS system can be given by
\begin{align*}
F_{\gamma_{o}^{BRS}}(\gamma)\to \frac{D^K2^{Kv}\gamma^{Kv}}{\overline\gamma^{KG_d}}.
\tag{25}\label{25}
\end{align*}
Therefore, the MGF of $\gamma_{o}^{BRS}$ is derived as
\begin{align*}
M_{\gamma_{o}^{BRS}}(s)= \frac{D^K2^{Kv}\Gamma(Kv+1)}{\overline\gamma^{KG_d}}s^{-Kv}.
\tag{26}\label{26}
\end{align*}
By inserting (26) into (20) and considering the BPSK modulation, the asymptotic SER of the BRS scheme can be formulated as
\begin{align*}
P_{e}^{BRS}\to \frac{D^K\Gamma(Kv+\frac{1}{2})2^{Kv}}{2\sqrt{\pi}\overline\gamma^{KG_d}}.
\tag{27}\label{27}
\end{align*}
From the above expression, it can be noticed that the achievable diversity order of the BRS scheme is maintained relative to that of the ARP scheme.
As such, we obtain the ratio
\begin{align*}
\frac{P_{e}^{BRS}}{P_{e}^{ARP}}= \frac{\Gamma(Kv+1)}{[\Gamma(v+1)]^K}\left(\frac{1}{K}\right)^{Kv}.
\tag{28}\label{28}
\end{align*}
From (28), the ratio is always smaller than 1 when $K\geq 2$. Therefore, we can conclude that based on the assumption of total power constraint, the BRS scheme is superior to the ARP system in terms of the SER performance. In fact, the ARP scheme tracks the fluctuations of all channels and assigns the best-condition channels to R and D. In addition, both S and R need transmit power to forward the information to the next node. If the total system power is limited to 1 and all sources and relays have the same transmit power, the transmit power at each node in the ARP scheme is $1/(2K)$. While for the BRS scheme, the transmit power of the selected S and R is $1/2$, respectively. Since the selective diversity gain is caused by the fluctuation of the channel fading, the larger transmission power at the source will cause the increase of the channel variation, thereby increasing the selective diversity gain. Similar work has also observed this phenomenon [48].

\begin{figure}[t]
    \centering
    \includegraphics[width=3.5in]{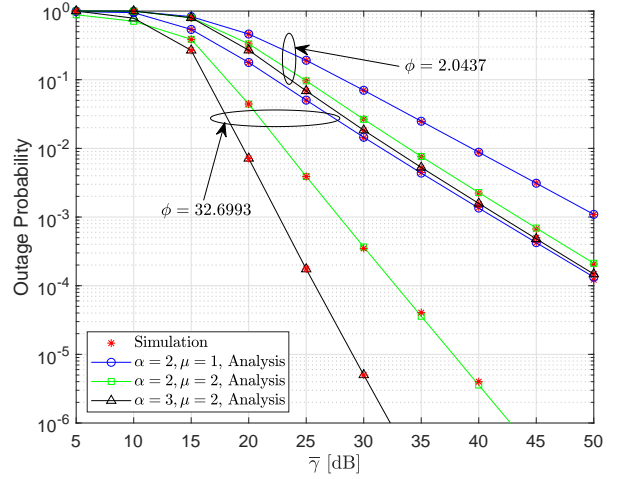}
    \caption{The OP of the dual-hop THz system versus $\overline\gamma$ with different
    fading parameters and pointing errors.}
\end{figure}

\section{Numerical Results and Discussions}
In this section, the Monte Carlo simulation is employed to verify the accuracy of the analytical results in this paper. Without loss of generality, a given threshold $\gamma_{th}=2$ dB and a fixed relay gain $C=1.7$ are considered. In addition, we assume that the average SNRs, fading parameters and pointing errors of two THz hops are equal, that is, $\overline\gamma_1=\overline\gamma_2=\overline\gamma$, $\alpha_1=\alpha_2=\alpha$, $\mu_1=\mu_2=\mu$, and $\phi_1=\phi_2=\phi$, except for Fig. 3 and Fig. 6. Unless otherwise is stated, we assume that the transmission distance of two links $d_1=d_2=d_o /2$, $d_o$ is the total transmit distance of the dual-hop THz system, the frequency and the antenna gain are respectively set as $f=300$ GHz and $G_t=G_r=55$ dBi, and the standard environment condition is considered, i.e., the relative humidity, atmosphere pressure and temperature are respectively $\varrho=50\%$, $p_a=101325$ Pa and $T=296$ K.

\begin{figure}[t]
    \centering
    \includegraphics[width=3.5in]{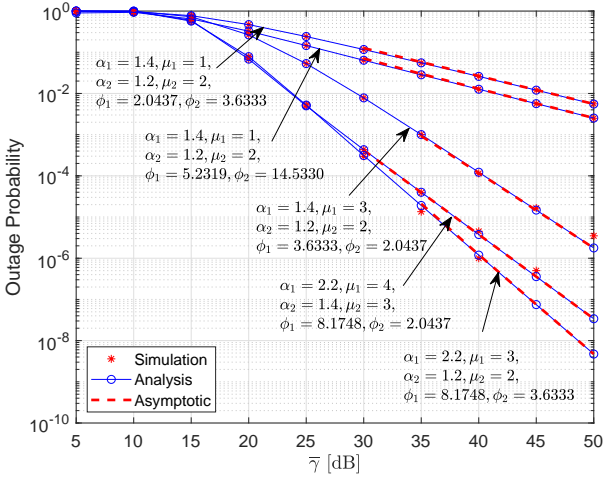}
    \caption{The OP of the dual-hop THz system for varying fading parameters and
    pointing errors along with the asymptotic results.}
\end{figure}

Fig. 2 presents the OP of the mixed dual-hop THz system versus $\overline\gamma$ with varying fading parameters and pointing errors. One can observe that as the values of fading parameters increase, the outage performance gets better since the larger values of $\alpha_1$, $\mu_1$, $\alpha_2$, and $\mu_2$ indicate the fading severity of THz links is reduced. Additionally, one can clearly see that pointing errors have a significant impact on the system outage performance. The pointing error parameter $\phi_i$ is related to $\sigma_{s,i}$, the lager the value of $\sigma_{s,i}$ is, the smaller the value of $\phi_i$ is, and higher impact of pointing errors is observed, which therefore leads to worse outage performance.

\begin{figure}[t]
    \centering
    \includegraphics[width=3.5in]{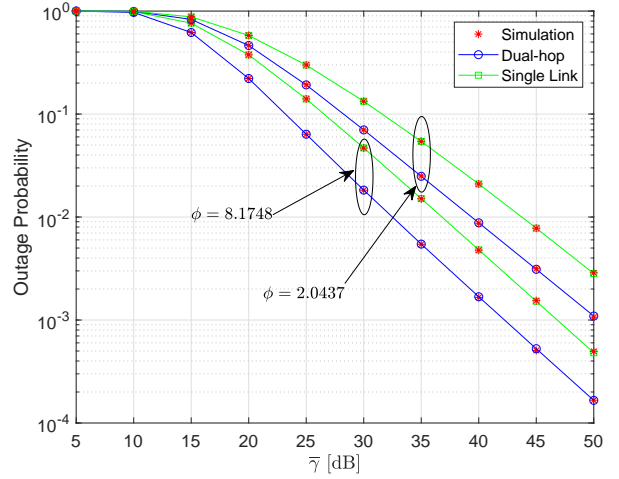}
    \caption{The OP of single THz and dual-hop THz links with different
    pointing errors for a total propagation distance of 100 m.}
\end{figure}

\begin{figure}[t]
    \centering
    \includegraphics[width=3.5in]{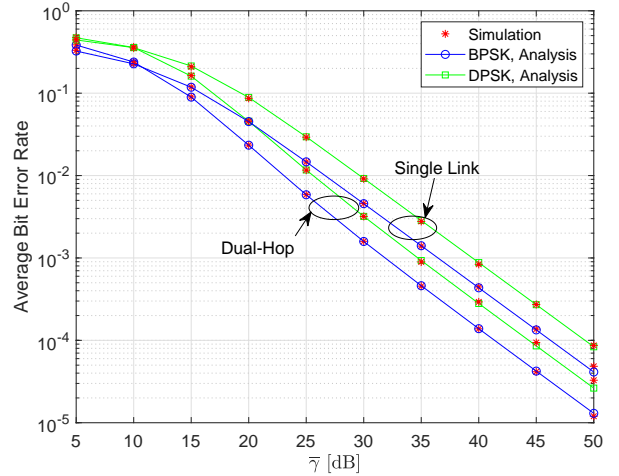}
    \caption{The average BER for different modulation schemes of single THz and dual-hop THz links with a total length of 80 m.}
\end{figure}

In order to verify the accuracy of the system's diversity order obtained from the asymptotic OP, Fig. 3 plots the OP versus $\overline\gamma$ with different fading parameters and pointing errors. As can be noticed that, the asymptotic results and the analytic expression of the OP have a tight fit at high SNRs. In addition, it can be observed that the curves have the same or different slopes as changing the values of $\alpha_1$, $\alpha_2$, $\mu_1$, $\mu_2$, $\phi_1$, and $\phi_2$, which is consistent with the diversity order $G_d=\min\left\{\frac{\phi_1}{2},\frac{\alpha_1\mu_1}{2},\phi_2,\alpha_2\mu_2\right\}$. Moreover, one can also notice that the OP decreases as the values of these fading parameters increase. Therefore, large diversity order can be obtained in the case of weak fading conditions and pointing errors, while strong fading and pointing errors result in a small diversity order.

\begin{figure}[t]
    \centering
    \includegraphics[width=3.5in]{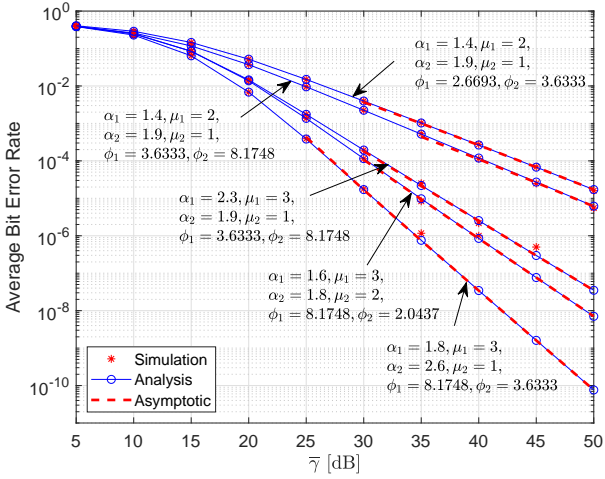}
    \caption{The average BER versus $\overline\gamma$ with different fading parameters and
    pointing errors along with the asymptotic results.}
\end{figure}

Fig. 4 shows the OP of single THz and dual-hop THz systems when the total transmission distance is $d_o=100$ m. The multipath fading parameters are set as $\alpha=2$ and $\mu=1$. For the dual-hop system, we consider $d_1=d_2=50$ m. One can see that under the same fading conditions and pointing errors, the OP of the dual-hop relaying THz system is lower than that of the single THz link. Moreover, it is observed in this figure that as the value of $\phi$ decreases, the impact of the pointing error becomes stronger, and therefore the outage performance becomes worse.

\begin{figure}[t]
    \centering
    \includegraphics[width=3.5in]{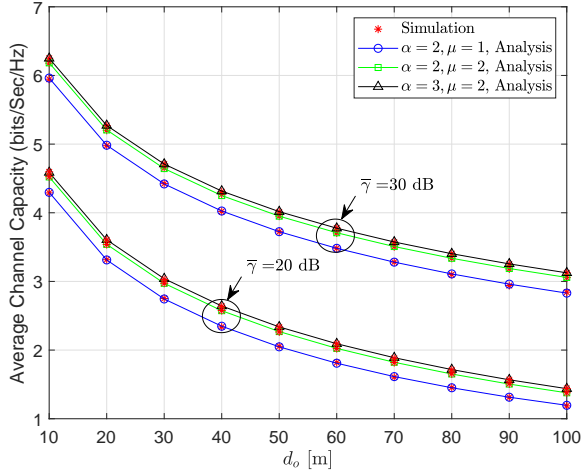}
    \caption{The ACC versus $d_o$ with different multipath fading parameters and average SNR $\overline\gamma$ when $\phi=3.6333$.}
\end{figure}

In Fig. 5, considering BPSK and DPSK modulation schemes, the average BER of the dual-hop system and the single link is presented. In this setup, the attenuation
parameters are set as $\alpha=2$, $\mu=3$, $\phi=2.0437$, and $d_o=80$ m. It can be clearly seen from this figure that the simulation results perfectly match numerical evaluated ones for all modulation schemes, and the dual-hop scheme has better error performance under two types of modulation schemes than the single THz link. Moreover, BPSK offers lower average BER than DPSK, as expected.

\begin{figure}[t]
    \centering
    \includegraphics[width=3.5in]{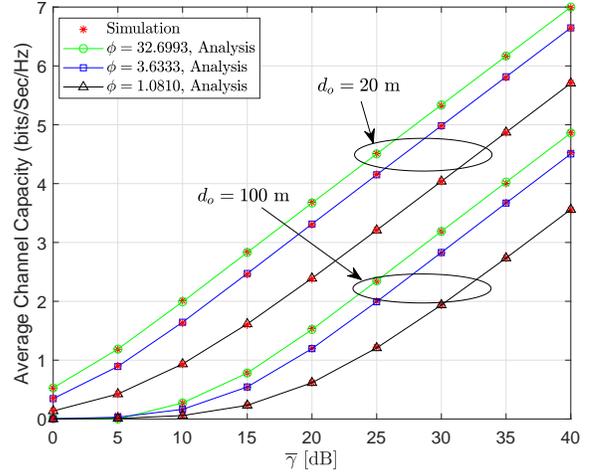}
    \caption{The ACC of the dual-hop THz system with different transmit distance and pointing errors.}
\end{figure}

Fig. 6 plots the average BER versus $\overline\gamma$ with varying fading parameters and pointing errors. In addition, the asymptotic average BER analysis given by (15) is illustrated in this figure. It is demonstrated that the proposed asymptotic result has excellent tightness and accuracy at high SNRs. Also, it can be clearly seen that the larger the value of $\alpha_1$ or $\mu_1$ is, the lower the average BER is. Furthermore, one can also observe that, with the increase of $\phi_1$ or $\phi_2$, the error performance of the dual-hop system gets better.

Fig. 7 illustrates the ACC versus the total propagation distance of dual-hop system $d_o$ with different multipath fading parameters and $\overline\gamma$ when $\phi=3.6333$. One can observe that the ACC decreases with the increase of $d_o$. The reason is that as the propagation distance increases, the system suffers from larger path loss and thus reduces the capacity. In addition, the ACC for different $\overline\gamma$ is also plotted. As expected, as $\overline\gamma$ increases, the average capacity increases. Also, one can see that the ACC gets larger value under the weaker fading.

\begin{figure}[t]
    \centering
    \includegraphics[width=3.5in]{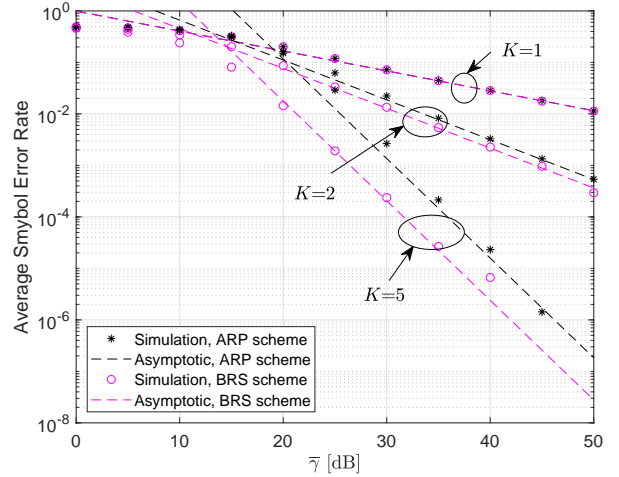}
    \caption{SER performance comparison: ARP versus BRS.}
\end{figure}

The curves of ACC versus $\overline\gamma$ for different propagation distances and pointing errors are plotted in Fig. 8. In this setup, the fading parameters are set as $\alpha=2,\mu=1$. One can clearly observe that the ACC is significantly affected by the transmission distance. As the value of $d_o$ increases, the path attenuation of the system increases, resulting in significant degradation of the system capacity performance. In addition, this figure shows the impact of pointing errors of the THz link on the ACC. For example, for a fixed $d_o$, it can be noticed that the ACC performance is getting better when the pointing error effect changes from strong to weak.

Fig. 9 depicts the SER performance comparison for ARP and BRS schemes. In this setup, the fading parameters are set as $\alpha_1=1.2$, $\mu_1=3$, $\alpha_2=1.3$, $\mu_2=2$, $\phi_1=1$, $\phi_2=3.6333$, and the total transmit distance is $d_o=100$ m. As can be clearly seen that the asymptotic results are match perfectly with the simulation results in high SNR regimes. For $K=1$, the system has only a single relay, therefore the SER performance of the BRS system is the same as that of the ARP case, as expected. In addition, one can observe that the two schemes have the same slopes, which implies that the diversity order of the BRS scheme is maintained with that of the ARP case , namely, $G_{d}^{Q}=K\min\left\{\frac{\phi_1}{2},\frac{\alpha_1\mu_1}{2},\phi_2,\alpha_2\mu_2\right\}$. Moreover, as the value of $K$ increases, the SER of both schemes is reduced, while the BRS scheme performs better than the ARP scheme.

\section{Conclusion}
In this paper, we have investigated the performance of dual-hop THz systems with fixed-gain AF relays. Taking the path loss, multipath fading and pointing errors into account, exact analytical expressions of the OP, average BER, and ACC were derived. Moreover, the accurate and tight asymptotic results of the OP and average BER were presented. Results demonstrated that the impact of multipath fading, pointing errors, and transmit distance significantly affects the performance of the dual-hop THz system. We also obtained that the diversity order of the mixed dual-hop THz system is $G_d=\min\left\{\frac{\phi_1}{2},\frac{\alpha_1\mu_1}{2},\phi_2,\alpha_2\mu_2\right\}$. The result shown that the diversity order is determined by multipath fading parameters and pointing errors. In addition, for the multi-relay system, the asymptotic SER expressions of both ARP and BRS schemes were derived. It was proven that the BRS fixed-gain AF relaying achieve the same diversity order $G_{d}^{Q}=K\min\left\{\frac{\phi_1}{2},\frac{\alpha_1\mu_1}{2},\phi_2,\alpha_2\mu_2\right\}$ as and lower SER than the ARP relaying.

\appendices
\section{CDF of The E2E SNR}
From (6), by defining $\mathcal{I}_1=\int_{0}^{\infty}F_{\gamma_2}\left(\frac{C\gamma}{x}\right)f_{\gamma_1}(x+\gamma)dx$, (6) can be rewritten as
\begin{align*}
F_{\gamma_o}(\gamma)=F_{\gamma_1}(\gamma)+\mathcal{I}_{1},
\tag{A.1}\label{A.1}
\end{align*}
where $F_{\gamma_1}(\gamma)$ is given in (5). Substituting (4) and (5) into $\mathcal{I}_1$ and consequently applying [33, Eqs. (9.301), (3.194.3) and (8.384.1)], we have
\begin{align*}
\mathcal{I}_1 &= \frac{2A_1A_2\overline\gamma_{1}^{-\frac{\phi_1}{2}}
\overline\gamma_{2}^{-\frac{\phi_2}{2}}C^{\frac{\phi_2}{2}}\gamma^{\frac{\phi_1}{2}}}{\alpha_2}\\
&\times\frac{1}{(2\pi i)^2}{\int\limits_{\ell_{1}}}{\int\limits_{\ell_{2}}}
\Gamma\left(\frac{\phi_2}{2}{-}\frac{\phi_1}{2}{+}\frac{\alpha_1}{2}t{-}\frac{\alpha_2}{2}s\right)\\
&\times\frac{\Gamma(s)\Gamma\left(s{+}\frac{\alpha_2\mu_2{-}\phi_2}{\alpha_2}\right)
\Gamma\left(1{-}\frac{\phi_2}{2}{+}\frac{\alpha_2}{2}s\right)\Gamma\left(\frac{\phi_2}{\alpha_2}{-}s\right)}
{\Gamma\left(s{+}1\right)\Gamma\left(1{+}\frac{\phi_2}{\alpha_2}{-}s\right)}\\
&\times\frac{\Gamma(t)\Gamma(t+\frac{\alpha_1\mu_1{-}\phi_1}{\alpha_1})}
{\Gamma(t{+}1)\Gamma\left(1{-}\frac{\phi_1}{2}{+}\frac{\alpha_1}{2}t\right)}\\&\times
\left(B_2\left(\frac{C}{\overline\gamma_2}\right)^{\frac{\alpha_2}{2}}\right)^{-s}
\left(B_1\left(\frac{\gamma}{\overline\gamma_{1}}\right)^{\frac{\alpha_1}{2}}\right)^{-t}dsdt,
\tag{A.2}\label{A.2}
\end{align*}
where $\mathcal{\ell}_1$ and $\mathcal{\ell}_2$ stand for the contours in the $s$-plane and the $t$-plane, respectively.
With the aid of (5) and (A.2) and by employing the definition of the BFHF, the CDF of the AF relaying system can be obtained as (7).

\section{Asymptotic CDF of The E2E SNR}
In this appendix, the asymptotic CDF of the considered system is derived. Specifically, when $\overline\gamma_1\to \infty$, the asymptotic expression of $F_{\gamma_1}(\gamma)$ is asymptotically derived by applying [40, Eq. (07.34.06.0040.01)] as
\begin{align*}
F_{\gamma_1}&(\gamma)\to \frac{2A_1\Gamma\left(\frac{\alpha_1\mu_1-\phi_1}{\alpha_i}\right)\Gamma\left(\frac{\phi_1}{\alpha_1}\right)
\gamma^{\frac{\phi_1}{2}}}{\alpha_1\Gamma\left(1+\frac{\phi_1}{\alpha_1}\right)}
\overline\gamma_{1}^{-\frac{\phi_1}{2}}\\
&+\frac{2A_1B_{1}^{\frac{\alpha_1\mu_1-\phi_1}{\alpha_1}}\Gamma\left(-\frac{\alpha_1\mu_1-\phi_1}{\alpha_1}\right)
\gamma^{\frac{\alpha_1\mu_1}{2}}}
{\alpha_1\mu_1\Gamma\left(1-\frac{\alpha_1\mu_1-\phi_1}{\alpha_1}\right)}\overline\gamma_{1}^{-\frac{\alpha_1\mu_1}{2}}.
\tag{B.1}\label{B.1}
\end{align*}
Moreover, relying on [37, Eq. (1.2)], $\mathcal{I}_1$ in (A.2) can be rewritten as
\begin{align*}
&\mathcal{I}_1 = \frac{2A_1A_2\overline\gamma_{1}^{-\frac{\phi_1}{2}}
\overline\gamma_{2}^{-\frac{\phi_2}{2}}C^{\frac{\phi_2}{2}}\gamma^{\frac{\phi_1}{2}}}{\alpha_2}\\
&\times\frac{1}{2\pi i}{\int\limits_{\ell_{1}}}\frac{\Gamma(s)\Gamma\left(s{+}\frac{\alpha_2\mu_2{-}\phi_2}{\alpha_2}\right)
\Gamma\left(1{-}\frac{\phi_2}{2}{+}\frac{\alpha_2}{2}s\right)\Gamma\left(\frac{\phi_2}{\alpha_2}{-}s\right)}
{\Gamma\left(s{+}1\right)\Gamma\left(1{+}\frac{\phi_2}{\alpha_2}{-}s\right)}\\
&\times\, {\mathrm{H}}_{2,3}^{3,0}\left [{{\frac{B_1\gamma^{\frac{\alpha_1}{2}}}{\overline\gamma_{1}^{\frac{\alpha_1}{2}}}}
\left |{ \begin{matrix} {(1,1), \left(1{-}\frac{\phi_1}{2},\frac{\alpha_1}{2}\right)}
\\ {\left(\frac{\phi_2}{2}{-}\frac{\phi_1}{2}{-}\frac{\alpha_2}{2}s,\frac{\alpha_1}{2}\right) (0,1) \left(\frac{\alpha_1\mu_1-\phi_1}{\alpha_1},1\right)}
\\ \end{matrix} }\right . }\right ]\!\\&\times
\left(B_2C^{\frac{\alpha_2}{2}}\overline\gamma_{2}^{-\frac{\alpha_2}{2}}\right)^{-s}ds.
\tag{B.2}\label{B.2}
\end{align*}
Assuming $\overline\gamma_1\to \infty$ and using [41, Eq. (1.8.4)] yields
\begin{align*}
\, {\mathrm{H}}_{2,3}^{3,0}&\left [{{\frac{B_1\gamma^{\frac{\alpha_1}{2}}}{\overline\gamma_{1}^{\frac{\alpha_1}{2}}}}
\left |{ \begin{matrix} {(1,1), \left(1{-}\frac{\phi_1}{2},\frac{\alpha_1}{2}\right)}
\\ {\left(\frac{\phi_2}{2}{-}\frac{\phi_1}{2}{-}\frac{\alpha_2}{2}s,\frac{\alpha_1}{2}\right) (0,1) \left(\frac{\alpha_1\mu_1-\phi_1}{\alpha_1},1\right)}
\\ \end{matrix} }\right . }\right ]\! \\
\to & \frac{2}{\alpha_1}\frac{\Gamma\left(\frac{\phi_1{-}\phi_2}{\alpha_1}+\frac{\alpha_2}{\alpha_1}s\right)
\Gamma\left(\mu_1{-}\frac{\phi_2}{\alpha_1}{+}\frac{\alpha_2}{\alpha_1}s\right)}
{\Gamma\left(1{-}\frac{\phi_2{-}\phi_1}{\alpha_1}{+}\frac{\alpha_2}{\alpha_1}s\right)
\Gamma\left(1{-}\frac{\phi_2}{2}+\frac{\alpha_2}{2}s\right)}\\
&\times \left(B_1\overline\gamma_{1}^{-\frac{\alpha_1}{2}}
\gamma^{\frac{\alpha_1}{2}}\right)^{\frac{\phi_2{-}\phi_1{-}\alpha_2s}{\alpha_1}}.
\tag{B.3}\label{B.3}
\end{align*}
Plugging (B.3) into (B.2) and using [37, Eq. (1.2)], and doing some algebraic manipulations, we obtain
\begin{align*}
\mathcal{I}_1 &\underset{\overline\gamma_{1} \to \infty}\approx \frac{4A_1A_2B_{1}^{\frac{\phi_2{-}\phi_1}{\alpha_1}}}{\alpha_1\alpha_2}\left(\frac{C\gamma}{\overline\gamma_{1}
\overline\gamma_{2}}\right)^{\frac{\phi_2}{2}}\\
&\times\, {\mathrm{H}}_{3,5}^{4,1}\left [{{B_{1}^{\frac{\alpha_2}{\alpha_1}}B_2
\left(\frac{C\gamma}{\overline\gamma_{1}\overline\gamma_{2}}\right)^{\frac{\alpha_2}{2}}}
\left |{ \begin{matrix} {\kappa_1}
\\ {\kappa_2}
\\ \end{matrix} }\right . }\right ]\!,
\tag{B.4}\label{B.4}
\end{align*}
where
$\kappa_1=\left\{\left(1{-}\frac{\phi_2}{\alpha_2},1\right) (1,1) \left(1{-}\frac{\phi_2{-}\phi_1}{\alpha_1},\frac{\alpha_2}{\alpha_1}\right)\right\}$,
$\kappa_2=\left\{(0,1) \left(\frac{\alpha_2\mu_2-\phi_2}{\alpha_2},1\right) \left(\frac{\phi_1{-}\phi_2}{2},\frac{\alpha_2}{\alpha_1}\right)
\left(\mu_1{-}\frac{\phi_2}{\alpha_1},\frac{\alpha_2}{\alpha_1}\right) \left(-\frac{\phi_2}{\alpha_2},1\right)\right\}$. Subsequently, assuming $\overline\gamma_2\to \infty$ and using [41, Eq. (1.8.4)], $\mathcal{I}_1$ can be further simplified. Finally, taking advantage of the asymptotic expression of $\mathcal{I}_1$ and (B.1), we get the asymptotic CDF of the considered system.

\section{PDF of The E2E SNR}
Plugging (4) in (10) and applying [33, Eq. (9.301)] and taking a series of transformation, we have
\begin{align*}
f_{\gamma_o}&(\gamma) = A_1A_2\overline\gamma_{1}^{-\frac{\phi_1}{2}}
\overline\gamma_{2}^{-\frac{\phi_2}{2}}C^{\frac{\phi_2}{2}}\gamma^{\frac{\phi_1}{2}{-}1}\\
&\times\frac{1}{(2\pi i)^2}{\int\limits_{\ell_{1}}}{\int\limits_{\ell_{2}}}
\Gamma\left(\frac{\phi_2}{2}{-}\frac{\phi_1}{2}{+}\frac{\alpha_1}{2}t{-}\frac{\alpha_2}{2}s\right)\\
&\times\frac{\Gamma(s)\Gamma\left(s{+}\frac{\alpha_2\mu_2{-}\phi_2}{\alpha_2}\right)
\Gamma\left({-}\frac{\phi_2}{2}{+}\frac{\alpha_2}{2}s\right)}
{\Gamma\left(s{+}1\right)}\\
&\times\frac{\Gamma(t)\Gamma(t+\frac{\alpha_1\mu_1{-}\phi_1}{\alpha_1})}
{\Gamma(t{+}1)\Gamma\left({-}\frac{\phi_1}{2}{+}\frac{\alpha_1}{2}t\right)}\\&\times
\left(B_2\left(\frac{C}{\overline\gamma_2}\right)^{\frac{\alpha_2}{2}}\right)^{-s}
\left(B_1\left(\frac{\gamma}{\overline\gamma_{1}}\right)^{\frac{\alpha_1}{2}}\right)^{-t}dsdt.
\tag{C.1}\label{C.1}
\end{align*}
By employing [37, Eqs. (2.56) and (2.57)], the expression of the PDF (C.1) can be rewritten by (11).
\section{Average BER of The E2E SNR}
For a single THz link, by inserting (5) into (13), the average BER $\overline{P}_{e1}$ can be derived by using
[40, Eqs. (07.34.03.0228.01) and (07.34.21.0012.01)] as
\begin{align*}
\overline{P}_{e1}&=\frac{A_1(q\overline\gamma_{1})^{-\frac{\phi_1}{2}}}{\Gamma(p)\alpha_1}\\ &\times
\, {\mathrm{H}}_{2,3}^{2,1}\left [{{\frac{B_1}{(q\overline\gamma_1)^{\frac{\alpha_1}{2}}}}
\left |{ \begin{matrix} {\left(1{-}p{-}\frac{\phi_1}{\alpha_1},\frac{\alpha_1}{2}\right) \left(1{-}\frac{\phi_1}{\alpha_1},1\right) (1,1)}
\\ {(0,1),\left(\frac{\alpha_1\mu_1-\phi_1}{\alpha_1},1\right), \left(-\frac{\phi_1}{\alpha_1},1\right)}
\\ \end{matrix} }\right . }\right ]\!.
\tag{D.1}\label{D.1}
\end{align*}
In addition, substituting (A.1) into (13) and applying [33, Eq. (3.326.2)] yields
\begin{align*}
\overline{P}_{e}& = \overline{P}_{e1}+\frac{A_1A_2\overline\gamma_{1}^{-\frac{\phi_1}{2}}
\overline\gamma_{2}^{-\frac{\phi_2}{2}}C^{\frac{\phi_2}{2}}q^{-\frac{\phi_1}{2}}}{\alpha_2\Gamma(p)}\\
&\times\frac{1}{(2\pi i)^2}{\int\limits_{\ell_{1}}}{\int\limits_{\ell_{2}}}
\Gamma\left(\frac{\phi_2}{2}{-}\frac{\phi_1}{2}{+}\frac{\alpha_1}{2}t{-}\frac{\alpha_2}{2}s\right)\\
&\times\frac{\Gamma(s)\Gamma\left(s{+}\frac{\alpha_2\mu_2{-}\phi_2}{\alpha_2}\right)
\Gamma\left(1{-}\frac{\phi_2}{2}{+}\frac{\alpha_2}{2}s\right)\Gamma\left(\frac{\phi_2}{\alpha_2}{-}s\right)}
{\Gamma\left(s{+}1\right)\Gamma\left(1{+}\frac{\phi_2}{\alpha_2}{-}s\right)}\\
&\times\frac{\Gamma(t)\Gamma(t+\frac{\alpha_1\mu_1{-}\phi_1}{\alpha_1})\Gamma\left(\frac{\phi_1}{2}{+}p{-}\frac{\alpha_1}{2}t\right)}
{\Gamma(t{+}1)\Gamma\left(1{-}\frac{\phi_1}{2}{+}\frac{\alpha_1}{2}t\right)}\\&\times
\left(B_2\left(\frac{C}{\overline\gamma_{2}}\right)^{\frac{\alpha_2}{2}}\right)^{-s}
\left(B_1\left(\frac{1}{\overline\gamma_{1}q}\right)^{\frac{\alpha_1}{2}}\right)^{-t}dsdt.
\tag{D.2}\label{D.2}
\end{align*}
Making full use of (D.1) and (D.2) and [37, Eq. (2.57)], the derived average BER of the dual-hop relaying system can be obtained as in (14).


\begin{thebibliography}{99}
\bibitem{1}
Z. Chen et al., ``A survey on terahertz communications,'' \emph{China Commun.},
vol. 16, no. 2, pp. 1-35, Feb. 2019.
\bibitem{2}
H. Elayan, O. Amin, B. Shihada, R. M. Shubair, and M.-S. Alouini, ``Terahertz band: The last piece of RF
spectrum puzzle for communication systems,'' \emph{IEEE Open J. Commun. Soc.}, vol. 1, 2020, pp. 1-32, Nov. 2020.
\bibitem{3}
K. Tekbiyik, A. R. Ekti, G. K. Kurt, A. Gorcin and H. Yanikomeroglu, ``A holistic investigation of terahertz
propagation and channel modeling toward vertical heterogeneous networks,'' \emph{IEEE Commun. Mag.},
vol. 58, no. 11, pp. 14-20, Nov. 2020.

\bibitem{4}
A. Afsharinejad, A. Davy, B. Jennings and C. Brennan, ``Performance analysis of plant monitoring nanosensor
networks at THz frequencies,'' \emph{IEEE Internet Things J.}, vol. 3, no. 1, pp. 59-69, Feb. 2016.
\bibitem{5}
C. Yi et al., ``Design and performance analysis of THz wireless communication systems for chip-to-chip and
personal area networks applications,'' \emph{IEEE J. Sel. Areas Commun.}, vol. 39, no. 6, pp. 1785-1796, June 2021.
\bibitem{6}
X. You, C. Wang, J. Huang, et al., ``Towards 6G wireless communication networks: vision, enabling technologies,
and new paradigm shifts,'' \emph{Sci. China Inf. Sci.}, vol. 64, no. 1, Nov. 2020.

\bibitem{7}
A. A. Boulogeorgos, E. N. Papasotiriou and A. Alexiou, ``Analytical performance assessment of THz wireless systems,''
\emph{IEEE Access}, vol. 7, pp. 11436-11453, 2019.
\bibitem{8}
K. Tekbiyik, A. R. Ekti, G. K. Kurt, A. Gorcin and H. Yanikomeroglu, ``A holistic investigation of terahertz
propagation and channel modeling toward vertical heterogeneous networks,'' \emph{IEEE Commun. Mag.},
vol. 58, no. 11, pp. 14-20, Nov. 2020.

\bibitem{9}
K. Tekbiyik et al., ``Statistical channel modeling for short range line-of-sight terahertz communication,'' in
\emph{Proc. IEEE 30th Annu. Int. Symp. Pers., Indoor Mobile Radio Commun. (PIMRC)}, Sep. 2019, pp. 1-5.
\bibitem{10}
F. Sheikh, M. El-Hadidy and T. Kaiser, ``Terahertz band: indoor ray-tracing channel model considering atmospheric
attenuation,'' \emph{Proc. IEEE AP-S/URSI}, Jul. 2015, pp. 1782-1783.
\bibitem{11}
F. Sheikh, N. Zarifeh, T. Kaiser, ``Terahertz band: Channel modelling for short-range wireless communications
in the spectral windows,'' \emph{IET Microw. Antennas Propag.}, vol. 10, no. 13, pp. 1435-1444, Oct. 2016.
\bibitem{12}
J. Kokkoniemi, J. Lehtom\"{a}ki, and M. Juntti, ``Simplified molecular absorption loss model for 275-400 gigahertz
frequency band,'' in \emph{Proc. 12th Eur. Conf. Antennas Propag. (EuCAP)}, London, U.K., Apr. 2018, pp. 1-5.
\bibitem{13}
A.-A. A. Boulogeorgos, E. N. Papasotiriou, J. Kokkoniemi, J. Lehtom\"{a}ki, A. Alexiou, and M. Juntti,
``Performance evaluation of THz wireless systems operating in 275-400 GHz band,'' in \emph{Proc. IEEE Veh.
Technol. Conf. (VTC)}, Jun. 2018, pp. 1-5.


\bibitem{14}
A. R. Ekti et al., ``Statistical modeling of propagation channels for terahertz band,'' in \emph{Proc. IEEE
Conf. Standards Commun. Netw. (CSCN)}, Helsinki, Finland, Sep. 2017, pp. 275-280.
\bibitem{15}
S. Priebe, C. Jastrow, M. Jacob, T. Kleine-Ostmann, T. Schrader, and T. Kurner, ``Channel and propagation
measurements at 300 GHz,'' \emph{IEEE Trans. Antennas Propag.}, vol. 59, no. 5, pp. 1688-1698, May 2011.
\bibitem{16}
M. D. Yacoub, ``The $\alpha-\mu$ distribution: A physical fading model for the stacy distribution,''
\emph{IEEE Trans. Veh. Technol.}, vol. 56, no. 1, pp. 27-34, Jan. 2007.

\bibitem{17}
A. A. Farid and S. Hranilovic, ``Outage capacity optimization for free-space optical links with pointing
errors,'' \emph{J. Lightw. Technol.}, vol. 25, no. 7, pp. 1702-1710, July 2007.
\bibitem{18}
W. Gappmair, ``Further results on the capacity of free-space optical channels
in turbulent atmosphere,'' \emph{IET Commun.,} vol. 5, no. 9, pp. 1262-1267, Jun. 2011.
\bibitem{19}
I. S. Ansari, F. Yilmaz, and M.-S. Alouini, ``Impact of pointing errors on the performance of mixed RF/FSO
dual-hop transmission systems,'' \emph{IEEE Wireless Commun. Lett.}, vol. 2, no. 3, pp. 351-354, June 2013.
\bibitem{20}
E. N. Papasotiriou, A.-A.-A. Boulogeorgos, and A. Alexiou, ``Performance analysis of THz
wireless systems in the presence of antenna misalignment and phase noise,''
\emph{IEEE Commun. Lett.}, vol. 24, no. 6, pp. 1211-1215, Jun. 2020.
\bibitem{21}
A.-A. A. Boulogeorgos and A. Alexiou, ``Error Analysis of Mixed THz-RF Wireless Systems,''
\emph{IEEE Commun. Lett.}, vol. 24, no. 2, pp. 277-281, Feb. 2020.

\bibitem{22}
E. Zedini, H. Soury, and M.-S Alouini, ``On the performance analysis of dual-hop mixed FSO/RF systems,''
\emph{IEEE Trans. Wireless Commun.}, vol. 15, no. 5, pp. 3679-3689, May 2016.
\bibitem{23}
B. Ashrafzadeh, E. Soleimani-Nasab, M. Kamandar, and M. Uysal, ``A framework on the performance analysis
of dual-hop mixed FSO-RF cooperative systems,'' \emph{IEEE Trans. Commun.},
vol. 67, no. 7, pp. 4939-4954, July 2019.
\bibitem{24}
Y. Zhang, J. Zhang, L. Yang, B. Ai, and M. Alouini, ``On the performance of dual-hop systems over mixed
FSO/mmwave fading channels,'' \emph{IEEE Open J. Commun. Soc.}, vol. 1, pp. 477-489, 2020.
\bibitem{25}
G. Xu and Z. Song, ``Performance analysis for mixed $\kappa-\mu$ fading and $\mathcal{M}$-distribution dual-hop
radio frequency/free space optical communication systems,''
\emph{IEEE Trans. Wireless Commun.}, vol. 1, pp. 477-489, 2020.
\bibitem{26}
P. Bhardwaj and S. M. Zafaruddin, ``Performance of dual-hop relaying for THz-RF wireless link,''
[Online]. Available: https://arxiv.org/abs/2012.13505, 2020.
\bibitem{27}
P. Bhardwaj and S. M. Zafaruddin, ``Performance analysis of dual-hop relaying for THz-RF wireless
link with asymmetrical fading,'' [Online]. Available: https://arxiv.org/abs/2103.08188, 2021.


\bibitem{28}
E. Zedini, H. Soury, and M. Alouini, ``Dual-hop FSO transmission systems over gamma-gamma turbulence
with pointing errors,'' \emph{IEEE Trans. Wireless Commun.}, vol. 16, no. 2, pp. 784-796, Feb. 2017.
\bibitem{29}
B. Ashrafzadeh, A. Zaimbashi, E. Soleimani-Nasab and M. Uysal, ``Unified performance analysis of multi-hop
FSO systems over double generalized gamma turbulence channels with pointing errors,"
\emph{IEEE Trans. Wireless Commun.}, vol. 19, no. 11, pp. 7732-7746, Nov. 2020.
\bibitem{30}
A. -A. A. Boulogeorgos and A. Alexiou, ``Outage probability analysis of THz relaying systems,''
in \emph{Proc. IEEE 31st Annu. Int. Symp. Pers., Indoor Mobile Radio Commun. (PIMRC)}, Oct. 2020, pp. 1-7.

\bibitem{31}
M. O. Hasna and M. S. Alouini, ``A performance study of dual-hop transmissions with fixed gain relays,''
\emph{IEEE Trans. Wireless Commun.}, vol. 3, no. 6, pp. 1963-1968, Nov. 2004.
\bibitem{32}
L. Yang, X. Yan, S. Li, D. B. da Costa and M. -S. Alouini, ``Performance analysis of dual-hop mixed PLC/RF
communication systems,'' \emph{IEEE Syst. J.}, Early Acccess, DOI: 10.1109/JSYST.2021.3088096.

\bibitem{33}
I. S. Gradshteyn and I. M. Ryzhic, \emph{Table of Integrals, Series, and Products.}, 7th ed. San Diego,
CA, USA: Academic Press, 2007.
\bibitem{34}
A. P. Prudnikov, Y. A. Brychkov, and O. I. Marichev, \emph{Integrals and Series: Vol. 3: More
Special Functions}. New York, NY, USA: CRC, 1992.

\bibitem{35}
S. Li, L. Yang, D. B. da Costa, J. Zhang and M.-S. Alouini, ``Performance analysis of mixed RF-UWOC
dual-hop transmission systems,'' \emph{IEEE Trans. Veh. Technol.}, vol. 69, no. 11,
pp. 14043-14048, Nov. 2020.
\bibitem{36}
S. Li, L. Yang, D. B. da Costa, and S. Yu, ``Performance analysis of UAV-based mixed RF-UWOC transmission systems,''
\emph{IEEE Trans. Commun.}, Early Access, DOI: 10.1109/TCOMM.2021.3076790.
\bibitem{37}
A. M. Mathai, R. K. Saxena, and H. J. Haubold, \emph{The H-Function: Theory and Applications}.
New York, NY, USA: Springer, 2010.
\bibitem{38}
K.~P. {Peppas}, ``A new formula for the average bit error probability of dual-hop amplify-and-forward
relaying systems over generalized shadowed fading channels,'' \emph{IEEE Wireless Commun. Lett.},
vol. 1, no. 2, pp. 85-88, Apr. 2012.
\bibitem{39}
E. Illi, F. El Bouanani, and F. Ayoub, ``A performance study of a hybrid 5G RF/FSO transmission system,''
in \emph{Proc. Int. Conf. Wireless Netw. Mobile Commun. (WINCOM)}, Nov. 2017, pp. 1-7.
\bibitem{40}
Wolfram., \emph{The Wolfram Functions Site}. [Online]. Available: http://functions.wolfram.com
\bibitem{41}
A. Kilbas and M. Saigo, \emph{H-Transforms: Theory and Applications (Analytical Method and
Special Function)}, 1st ed. CRC Press, 2004.
\bibitem{42}
Z. Wang and G. B. Giannakis, ``A simple and general parameterization quantifying performance
in fading channels,'' \emph{IEEE Trans. Commun.}, vol. 51, no. 8, pp. 1389-1398, Aug. 2003.
\bibitem{43}
F.~{Yilmaz} and M.-S~{Alouini}, ``Product of the powers of generalized nakagami-$m$ variates and performance
of cascaded fading channels,'' in \emph{in Proc. IEEE Global Telecommun. Conf.}, Nov./Dec. 2009, pp. 1-8.
\bibitem{44}
P. A. Anghel and M. Kaveh, ``Exact symbol error probability of a cooperative network in a Rayleigh-fading environment,''
\emph{IEEE Trans. Wireless Commun.}, vol. 3, no. 5, pp. 1416-1421, Sep. 2004.
\bibitem{45}
Y. Zhao, R. Adve, and T. J. Lim, ``Symbol error rate of selection amplify-and-forward relay systems,''
\emph{IEEE Commun. Lett.}, vol. 10, no. 11, pp. 757-759, Nov. 2006.
\bibitem{46}
M. K. Simon and M.-S. Alouini, \emph{Digital Communication Over Fading Channels: A Unified Approach to Performance Analysis}.
Hoboken, John Wiley \& Sons, Inc., 1st ed., 2000.
\bibitem{47}
J. Jung, S. Lee, H. Park, and I. Lee, ``Capacity and error probability analysis of diversity reception schemes
over generalized-K fading channels using a mixture Gamma distribution,'' \emph{IEEE Trans. Wireless Commun.},
vol. 13, no. 9, pp. 4721-4730, Sep. 2014.
\bibitem{48}
L. Yang and Q. T. Zhang, ``Performance analysis of MIMO relay wireless networks with orthogonal STBC,''
\emph{IEEE Trans. Veh. Technol.}, vol. 59, no. 7, pp. 3668-36674, Sep. 2010.

\end{thebibliography}
\end{document}